# Controlled domain wall pinning in ferromagnetic nanowire by nanoparticles' stray fields

V.L. Mironov[1,2], O.L. Ermolaeva[1], E.V. Skorohodov[1]

[1]Department of Magnetic Nanostructures,
Institute for Physics of Microstructures RAS, GSP-105, Nizhny Novgorod 603950, Russia

[2]Department of Physics and Nanoelectronics,
N. I. Lobachevsky State University, Nizhny Novgorod 603950, Russia

**Abstract**
We report the results of experimental investigations of controlled domain wall (DW) pinning in a ferromagnetic nanowire (NW) by stray fields of two uniformly magnetized bistable ferromagnetic nanoparticles (NPs) placed on either side of the NW and elongated parallel to the NW axis. We show by magnetic force microscopy measurements that DW pinning strength essentially depends on the orientation of the NP magnetic moments relative to the NW magnetization and can reach 20 mT. We also performed micromagnetic simulations confirming the influence of the magnetostatic interaction of the DW with the NP stray fields on the pinning strength. The possible realization of logic cell with switchable logic function is discussed.

## I. Introduction

The DW motion in ferromagnetic NWs is the subject of intensive investigations motivated by perspective applications in magnetic memory and magnetic logic devices [1, 2]. The computational processes in such devices are based on the controlled movement of DWs under the influence of external magnetic fields or spin-polarized currents [3-6]. In this regard, special researches are focused on the precise DW positioning in NW through the creation of local traps for controllable DW pinning / depinning. In particular, a number of studies is devoted to the DW pinning on the artificial defects of NW in the form of protrusions and recesses with various shapes [7-14]. These defects play the role of stationary traps, while the DW pinning / depinning control is performed by changing the magnitude and duration of the external magnetic field pulses. Other promising possibility of controllable DW pinning is the use of local magnetic fields generated by single-domain nanomagnets located near the NW [15-19]. The DW pinning in such systems is caused by magnetostatic interaction between DW and nanomagnets stray fields. In this case the pinning energy and depinning field strongly depend on the relative orientation of the magnetic moments in NW-NP system and can be easy changed by remagnetization in NP subsystem. In particular, the DW pinning based on the magnetostatic interaction of a NW with a ferromagnetic nanobar (NB) has been discussed in Refs. 15 and 16. It was shown that a tunable system of NBs located on one side of the NW enables the control of NB-DW interaction and in particular the realization of asymmetric interaction potentials. A combination of a NB and a trap was used recently for controllable DW pinning in nanoconstriction [17]. In this case the NB was located perpendicular to the NW near a notch. It has been demonstrated that the NB stray field substantially modifies the magnitude of the pinning energy, depending on the relative orientation of the DW and NB magnetic moments. The DW pinning in a NW with perpendicular magnetic anisotropy by in-plane magnetized NP positioned on top of the NW was demonstrated in [18]. On the other hand, recently we proposed alternative system consisting of a planar ferromagnetic NW and two elongated ferromagnetic NPs placed perpendicularly on either side of the NW [19]. It was shown that in this case the motion of DW along the NW is associated with overcoming the potential relief in the form of symmetric barriers and wells. In particular, it was demonstrated that the ratio of DW nucleation field and depinning fields in various configurations of NPs' magnetic moments allows one to realize a logic cell performing the "exclusive disjunction" logical operation (XOR). In current paper we present the investigations of controllable DW pinning in similar NW-NPs ferromagnetic system but with a different configuration of the NPs' stray fields, which leads to an asymmetric potential relief and possible implementation of another logic function.

## II. Experiment and methods

We investigated a permalloy ($Ni_{80}Fe_{20}$) planar system, which consists of a NW and two NPs placed one on either side of the NW. The scanning electron microscope (SEM) image of the NW-NP system is presented in Fig. 1.

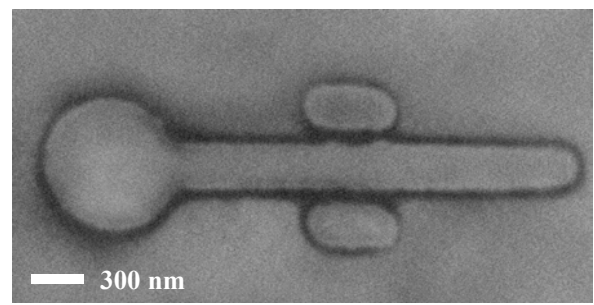

**Fig. 1.** SEM image of the NW-NP system.

The NW-NP system was fabricated by electron beam lithography and ion etching. Initially multilayer structure $Ni_{80}Fe_{20}$ (15 нм) / V (15 нм) / Cu (10 нм) was deposited onto a Si substrate using magnetron sputtering. Afterward the sample was covered by fullerene $C_{60}$ (80 nm), which was used as e-beam resist. The protective mask was formed in $C_{60}$ by exposure in the "ELPHY PLUS" system (based on the scanning electron microscope "SUPRA 50VP") with subsequent chemical treatment in an organic solvent. Afterward, the image was transferred to the Cu layer by $Ar^+$-ion etching and further to the V layer by plasma etching in Freon. At the final stage, the NW-NP system was fabricated in the ferromagnetic $Ni_{80}Fe_{20}$ layer by $Ar^+$-ion etching. The width of the NW was 270 nm, the NW length was about 2.8 μm; lateral dimensions of the NPs were $270 \times 500$ $nm^2$, the NP-NW separation was 70 nm. One end of the NW was fabricated in the form of circular disc for stimulated DW nucleation in this area [19-22]. The diameter of the nucleating part was 700 nm.

The magnetic states and the magnetization-reversal effects in the NW-NP system were studied using the magnetic force microscope (MFM) "Solver-HV", which is equipped with a dc electromagnet incorporated in a vibration-insulating platform (the maximal magnitude of the magnetic field is 0.1 T). The scanning probes were cobalt coated with a thickness of 30 nm. The tips were magnetized along the symmetry axes in a 1 T external magnetic field. The magnetic force microscope measurements were performed in the noncontact constant-height mode. The phase shift of cantilever oscillations under the gradient of the magnetic force was registered to obtain the MFM contrast [23].

Additionally, the DW structure and remagnetization effects in NW-NP system was studied by computer micromagnetic modeling using the standard object oriented micromagnetic framework (OOMMF) code [24]. The calculations were performed for the typical permalloy parameters: the saturation magnetization was $M_S = 8.6 \times 10^5$ A/m, the exchange stiffness was $A = 1.3 \times 10^{-11}$ J/m, and the damping constant was 0.5. We omitted magnetocrystalline anisotropy, assuming a polycrystalline structure of our samples. In calculations the NW was discretized into rectangular parallelepipeds with a square base of size $\delta = 2$ nm in the $x,y$ plane and height $h = 5$ nm.

## III. Results and discussion

We investigated DW pinning/depinning processes depending on the configuration of magnetic moments in NP subsystem. The experiments had the following protocol. Initially, the NW was magnetized from right to left (see Fig. 1) in an external magnetic field applied along the NW axis. Afterward we applied the reversed field (from left to right). In this case, reorientation of the magnetization in the circular pad and head-to-head DW formation were realized. The DW pinning and NW-NP system remagnetization were registered by MFM. The DW nucleation field for our system was $H_{nuc} = 13$ mT.

There are four different configurations of the NP magnetic moments, which are presented in Fig. 2.

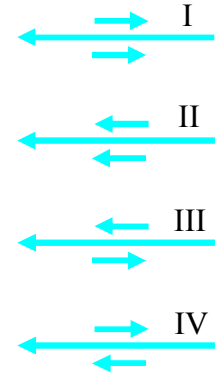

**Fig. 2.** The configurations of the NP magnetic moments.

First, we studied the NW magnetization reversal in the NW-NP system with magnetic moments corresponding to the configuration I. The different stages of the NW remagnetization experiment are presented in Fig. 3. The dark poles of MFM contrast correspond to south magnetic poles (while light poles correspond to north poles) in quasi-homogeneous distribution of the NP magnetization. The magnetic configuration of the system is shown schematically in the pictures by filled arrows. The initial state of the system was prepared as follows. First the NPs and NW were magnetized from right to left in strong external magnetic field. Afterward the inverse magnetic field with magnitude $H < H_{nuc}$ was applied and the direction of magnetization in both NPs was changed by local field of the magnetic force microscope probe. The technique of NP magnetization reversal is described in detail in Refs. 26 and 27. The MFM image of initial configuration I is presented in Fig. 3a. After the initial state preparation we applied gradually increasing external magnetic field opposite directed with respect to the NW magnetization. When $H$ exceeded 13 mT (DW nucleation field $H_{Nuc}$) we registered the appearance of an additional bright pole in the MFM image (indicated by the white arrow in Fig. 3b). As shown by micromagnetic simulation, the appearance of this additional pole can be explained by the formation of transverse head-to-head DW in the NW (corresponding magnetization distribution and model MFM image are presented in Fig. 4). The DW position was stable in external magnetic fields up to $H = 20$ mT, but when $H$ exceeded 20 mT (DW depinning field $H_{DI}$) we observed the remagnetization of the NW (see the change of the MFM contrast at the free NW end in Fig. 3c).

A different situation was observed in the magnetization reversal experiment with configuration II. The MFM images of consecutive stages of the magnetization reversal of the system for this case are shown in Fig. 5. Initial state was prepared by magnetizing from right to left in strong external magnetic field (Fig. 5a). After the applying of 13 mT reversed magnetic field we observed in the MFM image the appearance of additional bright pole in front of NPs (Fig. 5b). The micromagnetic simulation confirms that this additional pole can be explained by the formation of DW in the NW (the corresponding magnetization distribution and model MFM image shown in Fig. 6). After the applying of 15 mT field we observed the NW magnetization reversal accompanied by reorientation of NP magnetic moments (Fig. 5c).

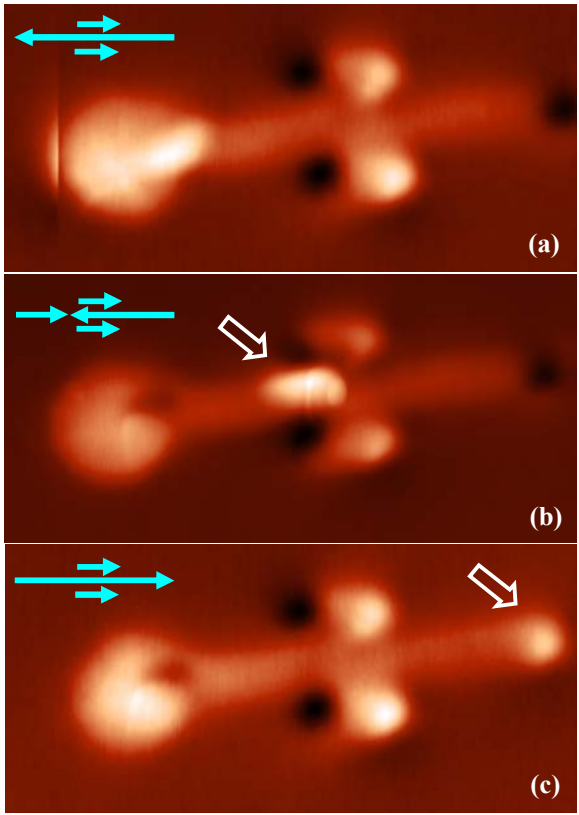

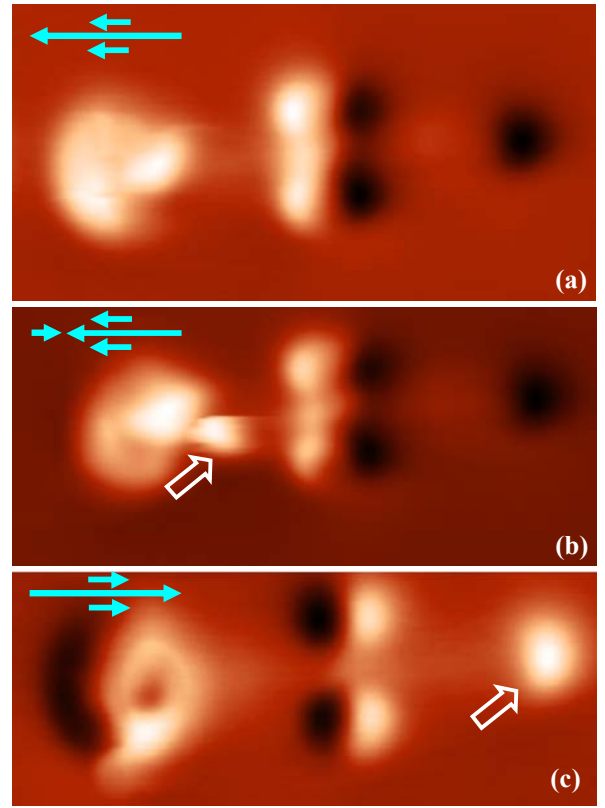

**Fig. 3.** The normalized MFM images of consecutive stages of magnetization reversal for NW-NP system with configuration I after the application of different external magnetic fields. The corresponding magnetic configuration of the system is presented schematically in the upper left corners of pictures by filled arrows. (a) Initial state with configuration I after previous magnetization. (b) The MFM image of NW-NP system after applying 13 mT external field. The DW is pinned directly between the NPs (the DW position is indicated by the white unfilled arrow). (c) The final stage after the remagnetization in 20 mT field. The changed NW pole is indicated by white unfilled arrow. The pad is in nonhomogeneous magnetic state.

**Fig. 5.** The normalized MFM images of consecutive stages of magnetization reversal for the NW-NP system with configuration II. The corresponding magnetic configuration of the system is presented schematically in the upper left corners of pictures by filled arrows. (a) Initial state with configuration I after previous magnetization. (b) The MFM image of NW-NP system in 13 mT external field. The DW is pinned near the NPs (the DW position is indicated by the white unfilled arrow). (c) The MFM image of NW-NP system after applying 15 mT external field (NW remagnetization is accompanied by reorientation of NP magnetic moments). The changed NW pole is indicated by white unfilled arrow.

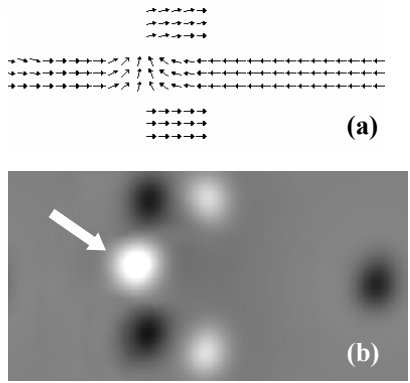

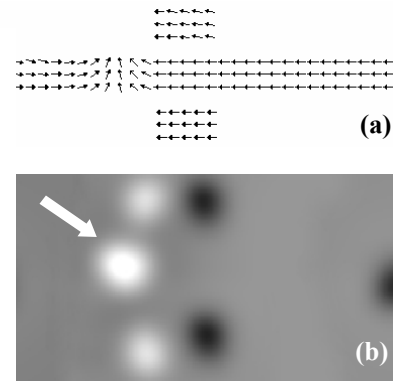

**Fig. 4.** (a) The model magnetization distribution in the NW-NP system (configuration I) corresponding to the DW pinning in the position before the NPs. (b) The model MFM contrast distribution from the NW-NP system (without nucleating pad) corresponding to the magnetization distribution shown in (a). The white arrow indicates the bright pole, which corresponds to the transverse head-to-head DW.

**Fig. 6.** (a) The model magnetization distribution in the NW-NP system corresponding to the DW pinning at the position before the NPs. (b) The corresponding model MFM contrast distribution from the NW-NP system (without nucleating pad). The white arrow indicates the bright pole, which corresponds to the head-to-head transverse DW.

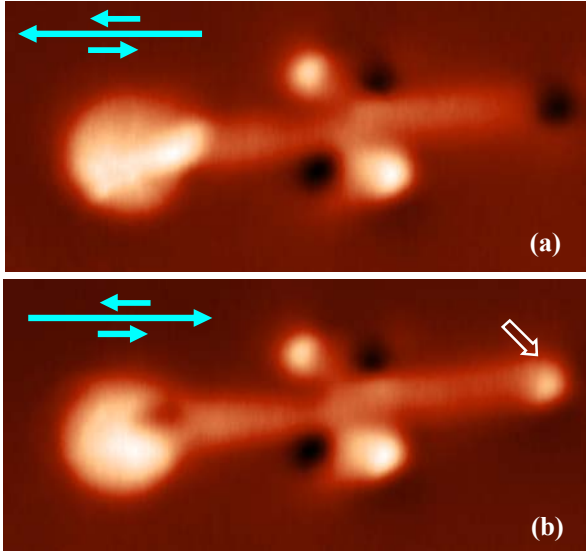

**Fig. 7.** The normalized MFM images of consecutive stages of magnetization reversal for the NW-NP system with configuration III. The corresponding magnetic configuration of the system is presented schematically in the upper left corners of pictures by filled arrows. (a) Initial state after previous magnetization. (b) The MFM image of NW-NP system after applying 13 mT external field. The changed NW pole is indicated by white unfilled arrow. The pad is in nonhomogeneous magnetic state.

In configurations III and IV the DW pinning effect was not observed experimentally. As an example, Fig. 7 shows the MFM images of successive stages of magnetization reversal in the system with magnetic configuration III. The initial state was prepared as follows. At the first stage NW and NPs were magnetized from right to left in a strong magnetic field. After that the direction of magnetization in one NP was changed by local field of the magnetic force microscope probe (Fig. 7a). After the initial state preparation we applied inversed external magnetic field opposite directed with respect to the NW magnetization. When $H$ exceeded 13 mT (DW nucleation field) we registered the NW remagnetization (Fig. 7b). This fact means that for configuration III the depinning field is smaller than the DW nucleation field ($H_{DIII} < H_{Nuc}$).

To estimate the pinning energy and depinning fields for different configurations of magnetic moments in the NP subsystem, we considered the dependence of the DW-NP interaction energy on DW position [19]. In the calculations we have neglected the distortion of the DW structure by NP fields and back influence of DW on the NP magnetization (the model of rigid DW and NPs [19, 25]).

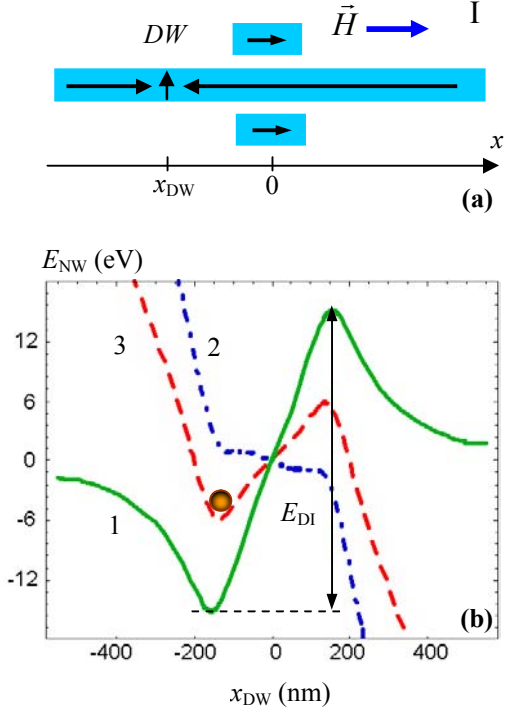

**Fig. 8.** (a) A configuration I of magnetization in the NW-NP system. (b) The profiles of potential energy $E_{NW}(x_{DW})$ for different external magnetic fields. The solid line 1 is the energy landscape at zero field. The dash-dot line 2 is for the depinning field $H_{DI} = 11$ mT. The dashed line 3 is for the intermediate field $0.5\,H_{DI}$. The DW pinning position is indicated schematically by the circle on curve 3.

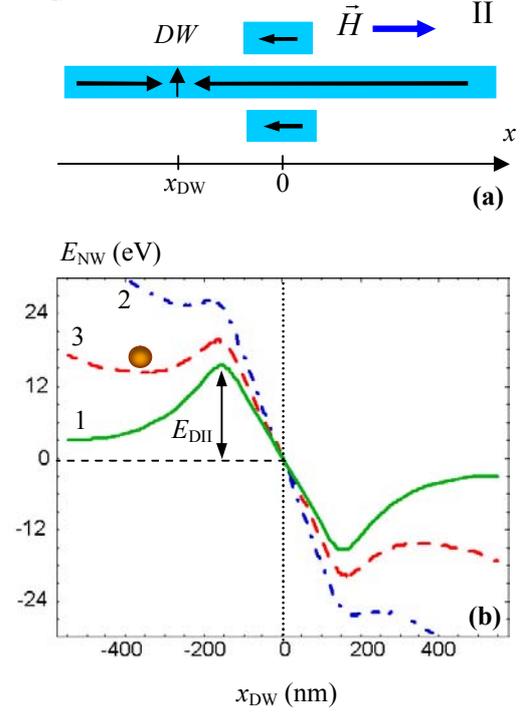

**Fig. 9.** (a) A configuration II of magnetization in the NW-NP system. (b) The profiles of potential energy $E_{NW}(x_{DW})$ for different external magnetic fields. The solid line 1 is the energy landscape at zero field. The dash-dot line 2 is for the depinning field $H_{DII} = 5.5$ mT. The dashed line 3 is for the intermediate field $0.5\,H_{DII}$. The DW pinning position is indicated schematically by the circle on curve 3.

The potential energy landscape $E_{NW}(x_{DW})$ in such system depends on the relative orientation of the magnetization in NW and NPs. The results of $E_{NW}(x_{DW})$ calculations for configuration I are shown in Fig. 8. The energy profile has the potential well and DW is pinned at the region before the NPs. The value of the energy well is defined mainly by the magnetostatic interaction of the NW magnetization outside the DW with the $x$ component of the NP field and does not depend on DW direction due to the system symmetry. The estimate of the activation energy $E_{DI}$ (in fact the value of right potential barrier) at zero field is 31.2 eV. In an external magnetic field the pinning barrier connected with the potential well is decreased (see curve 3 in Fig. 8b) and at 11 mT (depinning field $H_{DI}$) it vanishes completely (curve 2 in Fig. 8b) that corresponds to the DW depinning.

The energy landscape $E_{NW}(x_{DW})$ for configuration II is shown in Fig. 9b. The DW propagation is connected with overcoming the energy barrier in the region before NPs. The estimate of the energy barrier magnitude at zero field is $E_{DII}$ = 15.6 eV. In an external magnetic field the barrier is decreased (see curve 3 in Fig. 9b) and at 5.5 mT (depinning field $H_{DII}$) it vanishes completely (curve 2 in Fig. 9b).

The configurations III and IV correspond to the situation when the NPs' magnetic moments are opposite directed (Fig. 10a and Fig. 11a). Corresponding energy landscapes $E_{NW}(x_{DW})$ are presented in Fig. 10b and Fig. 11b. As seen from these figures the potential barriers for configurations III and IV are much lower than for configurations I and II. The potential barrier for the configuration III is $E_{DIII}$ = 4.8 eV and for configuration IV is $E_{DIV}$ = 2.4 eV. Corresponding depinning fields are $H_{DIII}$ = 2.3 mT and $H_{DIV}$ = 1.5 mT.

As is seen, the experimental results are in qualitative accordance with the theoretical estimations. However, there are considerable quantitative differences in the estimations of the fields $H_{DI}$ and $H_{DII}$. We believe that the high values of the depinning fields observed in the experiments can be connected with surface roughness, NW internal microcrystalline structure and changing of NP and DW magnetization structure due to strong DW-NP magnetostatic interaction. Note that the coercivity of NPs and DW depinning fields can be effectively tuned by the variation of NP shape and aspect ratio.

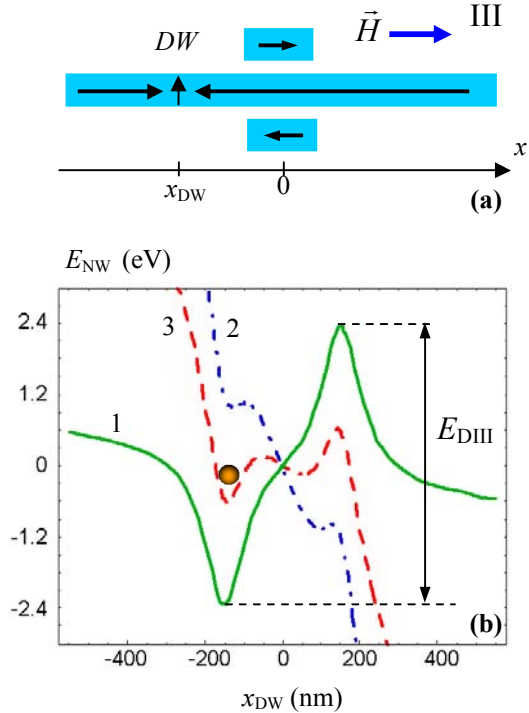

**Fig. 10.** (a) A configuration III of magnetization in the NW-NP system. (b) The profiles of potential energy $E_{NW}(x_{DW})$ for different external magnetic fields. The solid line 1 is the energy landscape at zero field. The dash-dot line 2 is for the depinning field $H_{DIII}$ = 2.3 mT. The dashed line 3 is for the intermediate field 0.5 $H_{DIII}$. The DW pinning position is indicated schematically by the circle on curve 3.

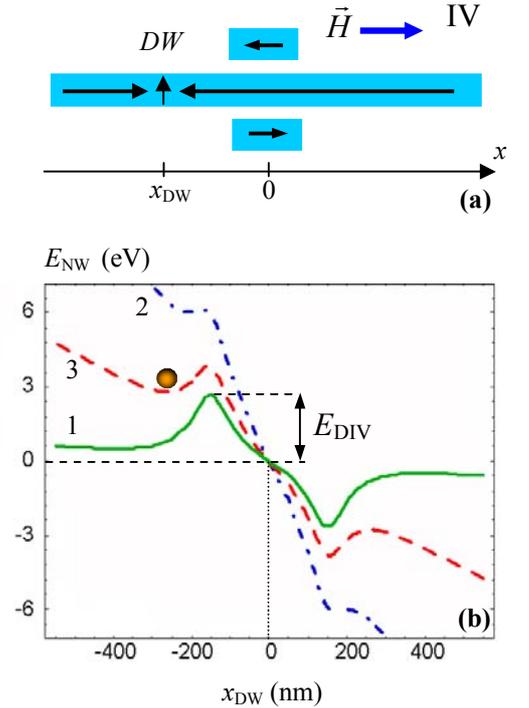

**Fig. 11.** (a) A configuration IV of magnetization in the NW-NP system. (b) The profiles of potential energy $E_{NW}(x_{DW})$ for different external magnetic fields. The solid line 1 is the energy landscape at zero field. The dash-dot line 2 is for the depinning field $H_{DIV}$ = 1.5 mT. The dashed line 3 is for the intermediate field 0.5 $H_{DIV}$. The DW pinning position is indicated schematically by the circle on curve 3.

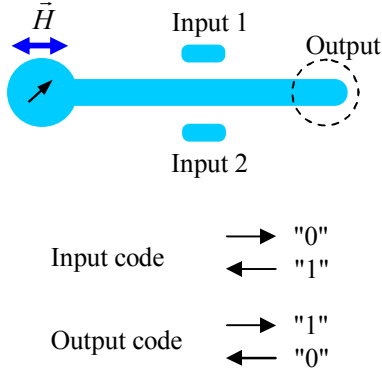

**Fig. 12.** Drawing of the schematic NW-NP logical cell and the coding of the input-output states.

Unlike the previously discussed system [19] the present NW-NP configuration has the asymmetric potential relief for DW motion. The relation between $H_{DI}$ and $H_{DII}$ enables the interesting possibility to realize a reconfigurable logic cell (LC). Schematic drawing of such LC is presented in Fig. 12. The direction of the NP magnetic moments plays a role of input information, and the magnetization direction at the end of the NW is output information. The operation algorithm for this LC is similar to the algorithm discussed in Ref. 19. The performing of logical operation consists of several stages. The first stage is the initialization process when the logical "0" is written in the NW by a uniform external magnetic field $H > H_{DI}$. Afterward the input information is written in the NPs by local magnetic fields, which can be created for instance with current busses or small magnetic coils. At the next step the reversed testing magnetic field $H_T$ is applied along the NW. Depending on the NP moments direction the DW either can be stopped at the gate, or can pass through the gate that leads to the NW remagnetization. At the final stage the output information is read. The magnetic state of the NW free end can be analyzed using the local magneto-optical Kerr effect or by means of local tunneling magnetoresistance measurement. Afterward the cycle of operations is repeated.

In contrast to the previously discussed system with NPs placed perpendicular to the NW [19], present system enables the realization of two different logical operation at the same logical element. If the testing magnetic field is less than the depinning field for the configuration II ($H_T < H_{DII}$), this LC performs the "exclusive disjunction" logical operation (so-called "XOR"). The correspondence between the input and output parameters for this case is shown in Tab. 1.

On the other hand, if the operating magnetic field is more than the depinning field for the configuration II, but less than the depinning field for configuration I ($H_{DII} < H_T < H_{DI}$), then this LC performs the "disjunction" logical operation (so colled "OR"). The correspondence between the input and output parameters for this case is shown in Tab. 2.

**Table 1.** The logical output states for all input states in the case of $H_T < H_{DII}$.

| Input 1 | Input 2 | Output |
|---------|---------|--------|
| 0 | 0 | 0 |
| 0 | 1 | 1 |
| 1 | 0 | 1 |
| 1 | 1 | 0 |

**Table 2.** The logical output states for all input states in the case of $H_{DII} < H_T < H_{DI}$.

| Input 1 | Input 2 | Output |
|---------|---------|--------|
| 0 | 0 | 0 |
| 0 | 1 | 1 |
| 1 | 0 | 1 |
| 1 | 1 | 1 |

**Summary**

Thus, we investigated the DW pinning effects in a system consisting of a ferromagnetic NW and two NPs placed on either side of the NW and elongated parallel to the NW axis. The theoretical estimations have showed that such NW-NP configuration has the asymmetric potential relief for DW motion. The DW pinning energy and corresponding depinning fields are strongly dependent on spatial configuration of magnetic moments in NP subsystem. The direct magnetic force microscopy measurements have showed that in dependence on relative orientation of magnetic moments in NW and NPs there are two variants of DW pinning connected with a potential well (configuration I) or a potential barrier (configuration II) caused by magnetostatic interaction between the DW magnetization in the NW and local NPs stray fields. For the $Ni_{80}Fe_{20}$ based NW-NPs system consisting of $270 \times 2800 \times 15$ nm nanowire with 700 nm in diameter nucleating part and $270 \times 500 \times 15$ nm nanoparticles gate (with 70 nm NW-NP separation) we registered the nucleating field $H_{Nuc} = 13$ mT and depinning fields $H_{DI} = 20$ mT and $H_{DII} = 15$ mT. When the magnetic moments of the NPs were set in configuration III or IV the DW pinning effects were not observed.

The observed relation between $H_{DI}$ and $H_{DII}$ potentially enables the realization of two different logical operations at the same logical element. If the testing magnetic field is less than the depinning field for the potential barrier ($H_T < H_{DII}$), this system performs the "XOR" operation. On the other side, if the testing magnetic field is more than the depinning field for the potential barrier, but less than the depinning field of the potential well ($H_{DII} < H_T < H_{DI}$), this system performs the "OR" operation. As the future developments, the different exotic compositions of nanowires and complex multi-particle gates based on monolayer and multilayer planar structures are very promising for the engineering of reconfigurable logic cells and DW memory systems.

**Acknowledgments**

The authors are very thankful to S. N. Vdovichev, A. Yu. Klimov and V. V. Rogov for assistance in sample preparation and A. A. Fraerman for very fruitful discussions. This work was supported by the Russian Foundation for Basic Research (Project No. 15-02-04462) and the programs of the Presidium of Russian Academy of Sciences. In experiments we used the equipment of the Center "Physics and technology of micro- and nanostructures".